\documentclass[10pt]{article}
\usepackage{a4}
\usepackage[final]{graphicx}    % Einbinden von EPS-Grafiken unterstuetzen
\usepackage[dvips]{epsfig}      % Einbinden von eps-Grafiken mit epsfig
\usepackage{psfrag}             % Ersetzen vom Text in EPS-Bildern durch LaTeX ermöglichen
\usepackage{latexsym}           % getting some symbols
\usepackage{exscale}            % adjust math-symbol size (especially for 12pt)
\usepackage{comment}            % provides the \begin{comment}\end{comment} environment.

\begin{document}
\pagestyle{plain}
\title {Feedback Medicine: Control Systems Concepts in \\Personalised, Predictive Medicine and Combinatorial Intervention\\}
\author{\\Peter Wellstead\footnote{peter.wellstead@nuim.ie}~\ Rick Middleton\footnote{Rick.Middleton@newcastle.edu.au}~\& Olaf Wolkenhauer\footnote{ow@informatik.uni-rostock.de}}

\maketitle
\setcounter{page}{1} \setlength{\footnotesep}{11pt}
\pagestyle{plain}

\section*{Introduction}

In its broadest definition, systems biology is the application of a `systems' way of
thinking about and doing cell biology \cite{wolkenhauer:05c}. By implication, this also invites us to consider
a systems approach in the context of medicine and the treatment of disease. In
particular, the idea that systems biology can form the basis of a personalised,
predictive medicine \cite{Hood04, RoyalSociety2005} will require that much closer
attention is paid to the analytic properties of the feedback loops which will be set up
by a personalised approach to healthcare. To emphasize the role that feedback theory
will play in understanding personalised medicine, we use the term \emph{feedback
medicine} to describe the issues outlined in these working notes.

More generally and beyond the personalised medicine debate, we advocate this `systems'
approach be extended to medicine as an important ingredient in future bio-medical
research. In this context, and in our interpretation, the systems approach pays close
attention to:

\begin{enumerate}
\item {Quantitative Modelling. Traditionally, due to the
difficulty of obtaining quantitative data, biological models have been qualitative and
descriptive rather than quantitative and predictive. In systems biology, we seek to
utilise emerging advances in sensing and analysis to generate and use quantitative
models. Apart from the obvious benefits of permitting predictive \emph{in silico}
modelling of biological system behaviour, there is also the potential for using more
powerful statistical tools in modelling, hypothesis testing, and parameter estimation to
be applied. However, for this potential to be realised data records will be required
that are much longer than currently available, collected using consistent and repeatable
measurement protocols, and with appropriate sampling rates. These three issues are
enormously important and, given the current status of bio-sensing and data collection,
they represent correspondingly enormous challenges to bio-sensing for systems biology.}

\item {System Dynamics. The transients, long term trends and
steady state behaviour of a biological system are all key
components in understanding the behaviour of a biological system.
Steady state behaviour alone is inadequate in many cases to give
the detailed knowledge of system response required for modern
medicine and surgery\footnote{The healing processes after surgical
intervention are all based to some degree or another on an assumed
biological mechanism for self-repair within organisms. By building
models of the dynamics of these mechanisms it is possible that
systems biologists may assist the development of surgical
practice.}.}

\item {Feedback and Interactions. The well known interconnections
of components in feedback, feedforward, nested systems and other
combinations are important elements in understanding the behaviour
of an entire system based on the interaction of it's constituent
parts \cite{wolkenhauer:05,wolkenhauer:05b}.}

\end{enumerate}

We note that this systems approach has its roots in the analysis
of dynamical systems, and thus differs substantially from the
bioinformatics perspective, which being grounded in computer
science focusses on data characterisation through pattern
recognition, correlation analysis, clustering and classification.
bioinformatics is an important way of developing quantitative
statements of the components in biology or processes that have
achieved steady state. However, one of the roles of system
dynamics and feedback theory in systems biology is to provide a
very distinct paradigm whereby we can \emph{analyse and
understand} the \emph{dynamical} behaviour of organisms. Thus
bioinformatics and systems biology perform complementary roles
that will increasingly overlap as their synergies develop.

In these notes we consider feedback and control systems concepts
applied to two important themes in medical systems biology -
personalised medicine \cite{Hood04} and combinatorial intervention
\cite{VanderGreef05}. In particular, we formulate a feedback
control interpretation for the administration of medicine, and
relate them to various forms of medical treatment. There are two
reasons for doing this. First, \emph{personalised medicine}
implies a tailoring of measurement, analysis of condition,
diagnosis and treatment to individual needs, that closely mirrors
the processes within a feedback control system. Second, the
notions of \emph{combinatorial} action \cite{Torrance00} in
treatment clearly indicate that the processes of medicine are, in
the terminology of control systems theory, multivariable and
coordinated in nature. For this reason we will use the term
\emph{coordinated}, in place of the phraseology connectivity and
combinatorial.

Personalised medicine implies a significant shift from what might
be termed population--based medical research, where the emphasis
is on determining how the population behaves on average. Instead,
this kind of information will be merged with information on how a
specific individual responds to various kinds of treatments. For
this to be viable, several ingredients may be needed such as:
\begin{itemize}
    \item {More regular monitoring, including self monitoring,
of important diagnostic indicators;}
    \item {Enhanced tools for learning appropriate individual information
    from time trends of individual diagnostic indicators;}
    \item {The need for treatment regimes and recipes that are widely applicable
    and give good clinical results for a variety of individuals with diverse steady state
    and dynamic responses to the same treatment\footnote{In the feedback control systems literature
    this is known as `robust control' where a feedback controller achieves good performance for
    a whole class of different `plants' to be controlled.}.}
\end{itemize}

Note that personalised medicine has often been discussed in
relation to \emph{pharmacogenetics} (see for example
\cite{RoyalSociety2005}). Our approach differs to this approach in
that we place a much higher emphasis on the role of dynamic
variables (such as concentrations of proteins and other variables)
than on the more static system parameters (e.g.\ genomic data)
typical of pharmacogenetics. Whilst some of the current
difficulties of implementing pharmacogenetics are shared by our
approach to personalised medicine, the role of dynamics in
prediction and regulation we believe to be an important ingredient
dictated by a systems biology approach.

In designing the enhanced tools and treatment regimes noted above
we believe that the tools of feedback theory are of potential
value. Biological processes are nonlinear, time-varying systems,
and in this form control theory is hard to apply. However, using
linearised stationary models, useful things can be said using
multivariable feedback system theory. In particular, multivariable
systems are known to have many special and subtle features which
have been exhaustively studied by control theorists, and there is
a large and informative literature on the subject (see e.g.\
\cite{SkogestadPostlethwaite2005}). Thus the objective of these
notes is to explore ways in which ideas from linear feedback
theory and practice might assist in the overall design of
personalised medicine and combinatorial intervention.

Our motivation in beginning this exercise now is the observation
that the technologies for automatically measuring the condition of
a patient are advancing rapidly. Thus it is both timely and
necessary to consider how feedback control ideas could help
diagnosis, treatment design and administration to be
systematically integrated with automated processes of measurement
within the \emph{personalised medicine} framework. A further hope
is that this systems approach may also clarify and give technical
basis for strategies adopted in some traditional approaches to
medicine. To reiterate our earlier remark we emphasise the central
role of feedback mechanisms in disease treatment as a whole by
refering to its study as \emph{feedback medicine}.

\section{The Systems Setup}

In this set of notes, we will not  discuss specific scientific
evidence for the model framework proposed. However, our (limited)
understanding of the literature is that the framework we use may
be of interest and is consistent with that implied in
\cite{Hood04, VanderGreef05}. The model framework we propose here
has two known limitations:
\begin{itemize}
\item {Our model here assumes no real ``intelligence" or
significant dynamics in disease effects, and therefore the results
here may well be inapplicable (at least in their current form) to
the treatment of infectious diseases (e.g.\ viral or bacterial).
In such cases, the disease itself may well show elements of
adaptive behaviour and this may be crucial in the design of
anti-bacterial or anti-viral therapies.}

\item{We do not directly address here the differences between
Homeostasis and recent work on Allostasis (e.g.\
\cite{Sterling2004}). We note that the difference in these
viewpoints, whilst potentially important from a medical viewpoint,
does not significantly alter the discussions contained here.
Homeostasis can be considered, within a control systems framework,
as a \emph{regulation problem}, whereas Allostasis resembles the
\emph{servo} or \emph{setpoint tracking problem}\footnote{The
function of a regulatory feedback controller is to maintain a
system output at a specified level (e.g Homeostasis), while a
tracking or servo feedback controller is designed such that the
output tracks changes in some reference variable which is defined
externally to the feedback loop (e.g.\ Allostasis).}. }
\end{itemize}

\begin{figure}[htbp]
\begin{center}
\includegraphics[width=5.5in]{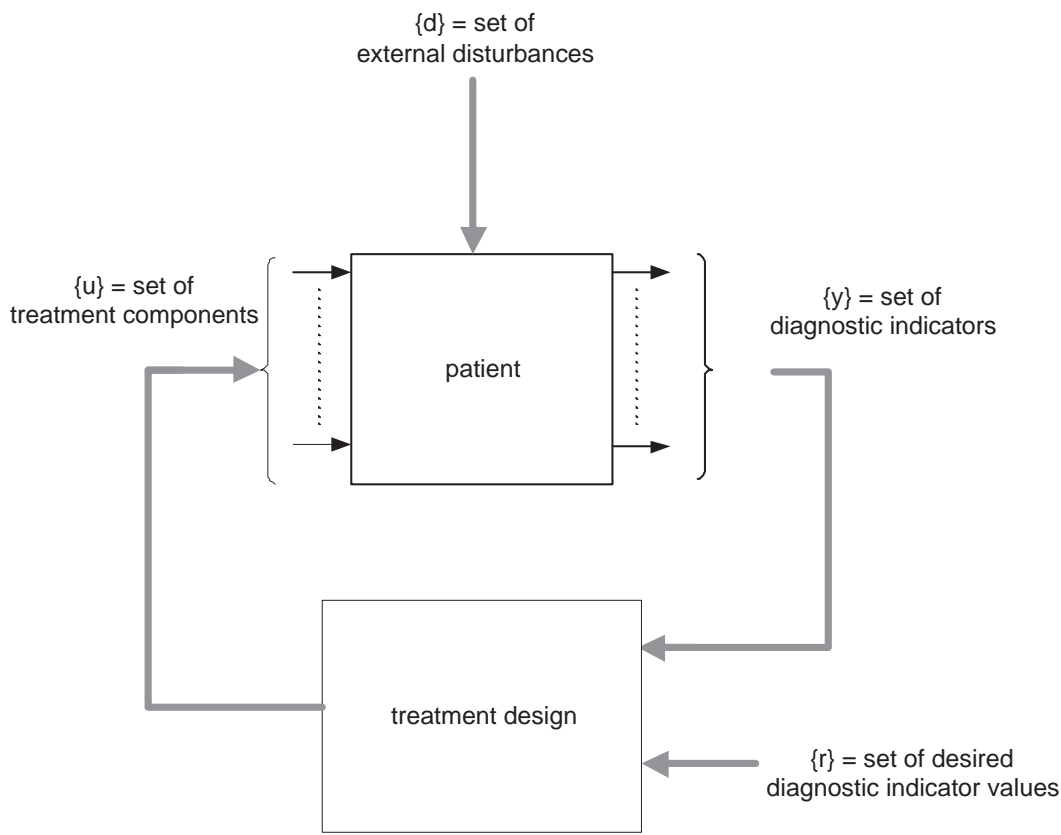}
\caption[A feedback control interpretation for treatment of
disease.]{A feedback control interpretation for treatment of
disease.} \label{fig:feedbackmedicine1}
\end{center}
\end{figure}

In a general form, Figure \ref{fig:feedbackmedicine1} shows a
feedback systems representation of the disease treatment process
in a way that mimics the actions of a physician. In context of
this figure treatment design includes diagnosis, selection of
treatments, and potentially the design of individual treatments.
Before going into the specifics of each of these, it will be
useful to introduce some systems terminology and seek out
biological parallels.

\subsection{Dynamic Variable Definitions}
\label{subsec:DynamicVars}

\textbf{Diagnostic Indicators (y):} In Figure
\ref{fig:feedbackmedicine1} the state of health of a patient is
determined indirectly by a set of `diagnostic indicators',
\{\textbf{y}\} which are a set of measurements, that change with
time, and that when taken together indicate the patient's response
to some treatment and/or their state of health. In control systems
terminology, \textbf{y} is a vector valued function of time, often
referred to as the \emph{output}.

\textbf{Reference levels for Diagnostic Indicators (r):} In the
block `treatment design' the diagnostic indicators \{\textbf{y}\}
are compared with the desired (or \emph{reference}) diagnostic
indicators denoted by \{\textbf{r}\}. Note that it would not be
uncommon for there to be a range of acceptable values for the
diagnostic indicators, in other words, perfect tracking of
\textbf{y} to \textbf{r} is not required. \textbf{r} is often
called an `exogenous' input, since this variable is typically
predefined based on relevant medical research.

Each member of the set \{\textbf{r}\} will comprise a set of
permissable values corresponding to the range of values that might
be anticipated in a healthy person of the class being considered.
Currently, these consist of a set of indicators that are learnt by
physicians as part of their training. No one physician has a
mastery of all possible indicators and the set of known indicators
is incomplete -- being constrained by measurement technology and
the boundaries of medical knowledge.

\textbf{Treatment Components (u):} The treatment components are
largely at the disposal of the physician and patient where timing
and dosage of a range of different treatments (e.g.\ drugs) can be
applied. In control systems terminology, \textbf{u} denotes the
vector of inputs (as a function of time), which need to be
specified as a function of the historical and current values of
\textbf{r} and \textbf{y}. An expanded discussion of how we think
of treatment components is given later in Subsection
\ref{subsec:ActuationInteractn}.

\textbf{Disturbances (d):} The disturbances denote another set of
variables \{\textbf{d}\}, often not measurable, that are not under
direct control of the patient or doctor, but which have a
significant effect on the patient, and the diagnostic indicators.
For example, stress, food, toxins might all be considered as
possible system disturbances. In control systems terminology,
these are often described as `exogenous disturbances'.

\subsection{Internal Variables: States and Model Parameters}
The variables described above in Subsection
\ref{subsec:DynamicVars} are all variables that are either
external to the patient, or can be measured externally. Usually
these external variables provide only an incomplete indication of
the underlying physiological condition of the patient. To complete
the picture and to allow prediction of future output values, two
further sets of variables are used in dynamic systems theory. The
first of these are \emph{Model Parameters} that show little or no
dynamic variation, but are needed to define in a quantitative way
the model behaviour. Examples of model parameters in biological
systems are things such as the relevant components of an
individual's genomic data, reaction rate constants, decay rates,
diffusion constants etc. Some of these may be difficult to
determine in biological systems from first principles, with only
qualitative information being available. Where they can be
quantified however it may be that real time measurements together
with system identification techniques can be used to find
numerical values for model parameters.

The second set of important internal variables is the state
vector. In dynamical systems theory the state \{\textbf{x}\}
completely and independently describes the current status of a
system and when taken with an appropriate mathematical model
(including values for all parameters) of the system, and
specification of all future inputs and disturbances, completely
specifies the future behaviour of the system \cite{MacFarlane70}.
In other words the state \{\textbf{x}\} is central to the ability
to predict a system's behaviour -- a point that has great
relevance to in our interpretation of predictive medicine as
outlined in \cite{Hood04}.

The concept of system state is extremely powerful and accordingly
it occupies a central role in the theory and analysis of dynamical
systems. The ways of determining a complete and independent set of
states for systems made up of inanimate objects is well understood
\cite{Wellstead79}, but this is not true in the case  of living
organisms. An important question in the systems approach to
biology and medicine is therefore: Does an analogous concept of
state exist in biology and medicine? We suggest that if the
central dogma of biology holds true, then the set of
concentrations of all proteins synthesised by an organism is a
natural candidate for the state \{\textbf{x}\} of that organism.

\begin{figure}[htbp]
\begin{center}
\includegraphics[width=4.5in]{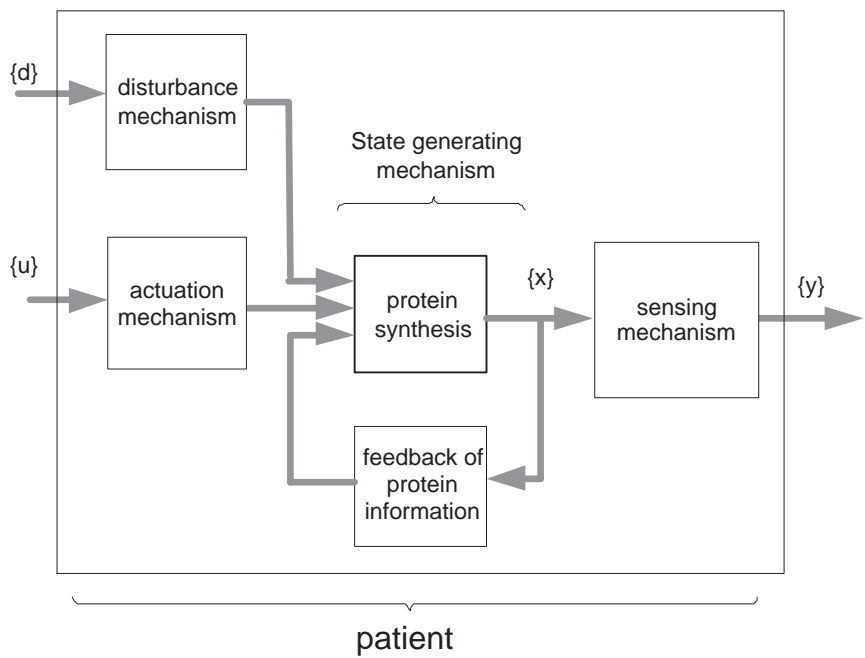}
\caption[Input, output and state generating mechanisms.]{Input,
output and state generating mechanisms.}
\label{fig:feedbackmedicine1b}
\end{center}
\end{figure}

We can put this in diagrammatic form in Figure
\ref{fig:feedbackmedicine1b} where we indicate how the patient
might be represented, in general terms, in a feedback systems
formulation. While this figure is meant only to be illustrative it
shows how the different aspects of medicine might be approached
from a systems viewpoint. First note that the treatment set
\{\textbf{u}\} corresponds in feedback control terms to a set of
actuation signals. By analogy we suppose that the treatment
reaches the central state generating mechanism -- which produces
the set of proteins -- via a form of biological actuation
mechanism. The task of the treatment designer is to understand the
actuation mechanism sufficiently to ensure that the treatment
produces the desired affect on the patient state. Typically, this
desired affect is to correct a particular part of the state which
has deviated significantly from it's desired concentration,
without disturbing other parts of the state that are already at
appropriate levels.

Note also the role of the block labelled `sensing mechanism'. This
produces measurements that form the diagnostic indicators, or
outputs, as some combination of the states or subset of the
states. In physical systems this would be achieved with set of
physical measuring devices, such as voltmeters, accelerometers,
pressure sensors or other tranducers. In medicine the set
\{\textbf{y}\} is a set of measureable variables which offer an
outward expression of the changes in system state. Given that the
range of measurable variables may be limited in a medical
situation, it is the task of the physician or diagnostician to
ensure that they select a set \{\textbf{y}\} that can be used to
infer the relevant values of \{\textbf{x}\}.

\subsection{Soft Sensing and State Observation}

In cases where the states are needed, but are not all directly
measurable, then algorithms called state observers/estimators are
used to reconstruct/estimate the actual states
\cite{MiddletonGoodwin90}. These observers are sometimes called
`soft sensors' since they use mathematical representations and a
time history of sensed variables to estimate or observe other
state variables. Design of such observers typically requires
mathematical models of the process in hand, including parameter
values, and in medicine this is not generally feasible. Hence we
assume that only the set of diagnostic indicators \{\textbf{y}\}
is available, but at the same time note the extreme importance of
dynamical models in eliciting information concerning system
behaviour when the corresponding state is not measurable
\footnote{An interesting question here might be, given the large
number of proteins, what are the relevant subsets for particular
conditions. To a degree, proteomics, aided by techniques from
graph theory, attempts to address this issue. The important issue
here is how to determine the sub--set of proteins (states)
required a personalised feedback medicine approach to a particular
disease.}.

In automated personalised medicine the aspiration is that the
diagnostic indicators could be a comprehensive list of
concentrations of bio--molecules in biofluids such as blood,
urine, mucus, etc\footnote{Metabonomics is a term used to describe
the bioinformatic analysis of biofluids, see for example
http://www.metabometrix.com}. Ideally, if the concentrations of
all bodily proteins could be measured, then diagnosis and
treatment design could be made on the basis of complete knowledge
of the patient's state. This complete state knowledge implies a
requirement for the time histories of relevant proteins for
different disease conditions. If this were to be possible then the
predictive capabilities of models of disease dynamics would make a
potent diagnostic tool. An important caveat in this respect is
that dynamical models in technical systems are much simpler than
in biology and physiology. However, given time, effort and
resource it is possible to model organ function at a usefully
detailed level, e.g.\ the virtual heart project \cite{Noble02} and
elements of the human physiome project \cite{HunterBorg03}.

\subsection{The Actuation Process and Interaction}\label{subsec:ActuationInteractn}
As noted previously, the process of administering medication is
the biological analogue of actuation in a technological control
system. It is relevant to note here are that in technological
systems we can usually be absolutely precise in applying a control
signal to a system in a desired way. For example, controlling the
current supplied to an electrical motor produces precisely known
changes in motor torque -- and nothing more. On the other hand
medication will in general influence many body components in
addition to the disease target. For example, pharmacologists speak
of `dirty' drugs whereas the control systems analyst would talk of
highly interacting or strongly coupled actuation.  This strong
coupling, or multivariable, effect of a treatment upon the states
is at the heart of the combinatorial treatment concept, and is
this an important issue in feedback medicine. Specifically,
multivariable control theory has special procedures for dealing
with interaction and reducing its effects on an overall system.
Thus it may be possible for multivariable feedback theory ideas to
help understand how interacting treatments behave and coordinate
them in some manner. We note in passing the duality between the
actuation mechanism and the measurement mechanism. With the dual
roles of the person who designs a medical treatment, who must
understand the actuation process, and the diagnostician, who must
understand the measurement process.

\section{Issues in Conventional Feedback Medicine}

\begin{figure}[htbp]
\begin{center}
\includegraphics[width=5.5in]{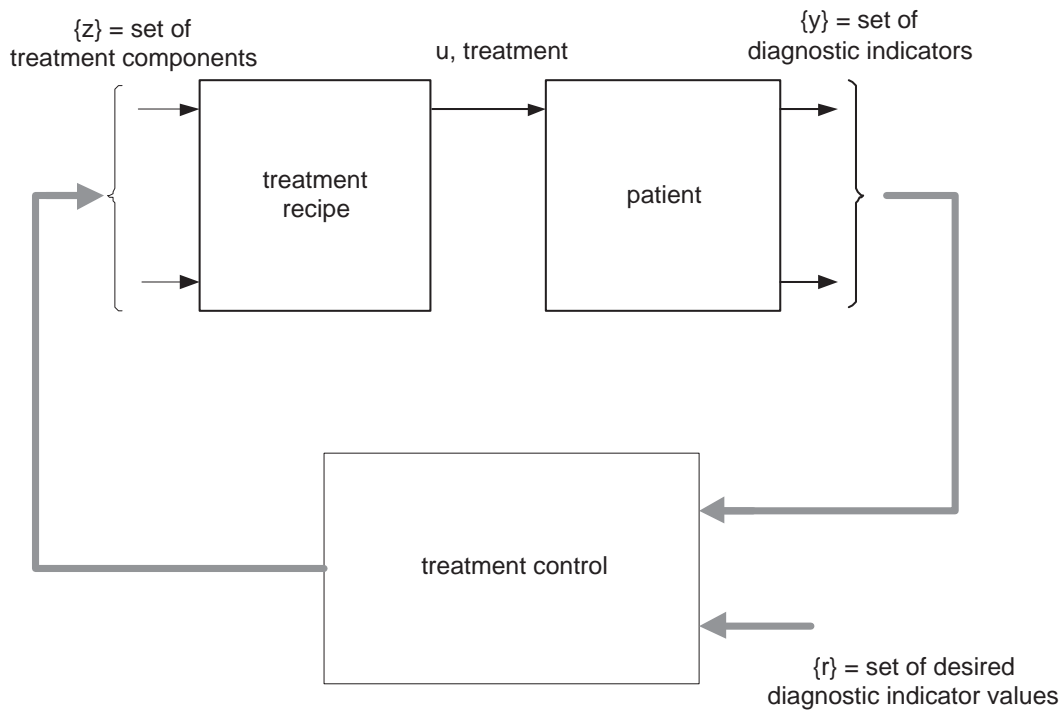}
\caption[A feedback control interpretation for treatment of disease: 2]{A feedback
control interpretation for treatment of disease: 2.} \label{fig:feedbackmedicine2}
\end{center}
\end{figure}

If we assume that the set, \{\textbf{y}\} is comprehensive in the
sense that it covers all relevant conditions, and that a (correct)
diagnosis has been made, then a possible control block diagram for
\emph{feedback medicine} is shown in Figure
\ref{fig:feedbackmedicine2}. Here we assume that one treatment is
prepared from a recipe of components from the set \{\textbf{z}\}
specified by the treatment controller. Thus:

\begin{equation}
u=\sum_{i=1}^n b_{i}z_{i} \label{eqn:recipe}
\end{equation}

In the unrealistic linear case with $n$ independent outputs
\{\textbf{y}\} and n independent components to the treatment
\{\textbf{z}\}, and laying dynamical response issues to one side
for the moment, it is possible to select the combination of the
inputs in equation \ref{eqn:recipe} to form a medication
\{\textbf{u}\} which achieves a desired diagnostic state
\{\textbf{r}\} `open loop'. That is to say with one treatment step
and without the further need for feedback in the form of continued
consultation with a physician (the treatment controller). In
general however the patient's response to treatment is nonlinear
with unknown and time varying dynamics such that the idea of a
general analysis of feedback medicine is at this stage illusory.
There are however some general points from control theory that can
be made with for configuration shown in Figure
\ref{fig:feedbackmedicine2}, and in the following we will attempt
to explore them.

\subsection{Fixed Remedies}
If only proprietary drugs and medicines are available, then the
weightings on the input treatments in equation \ref{eqn:recipe}
can not be changed by the physician/controller, in other words
\{\textbf{u}\} is largely fixed by the drug designer \textit{a
priori} with only few degrees of freedom (e.g.\ frequency and
dosage) available to the physician. In this context, a proprietary
drug is similar to a controller with fixed parameters. In
medicine, as in technological feedback systems, important degrees
of freedom are removed from the treatment controller by fixing the
parameters. They can cause poor dynamical performance and prevent
the required therapeutic state being achieved. The fact that the
desired therapeutic state is a set of acceptable bands may
overcome this. Specifically, it may be possible for an incorrect
treatment to move the vector of diagnostic indicators into an
acceptable final (steady state) region. The question of a poor
dynamical response to a treatment is however an issue that that is
rarely if ever discussed in medicine or biology. More
specifically, there seems to be little recognition of the
importance of the transient component of a patient's response to
medication. We will return to this point in Section
\ref{section:Personalised FM} of these notes.

\subsection{Switching Treatments}

\begin{figure}[htbp]
\begin{center}
\includegraphics[width=5.5in]{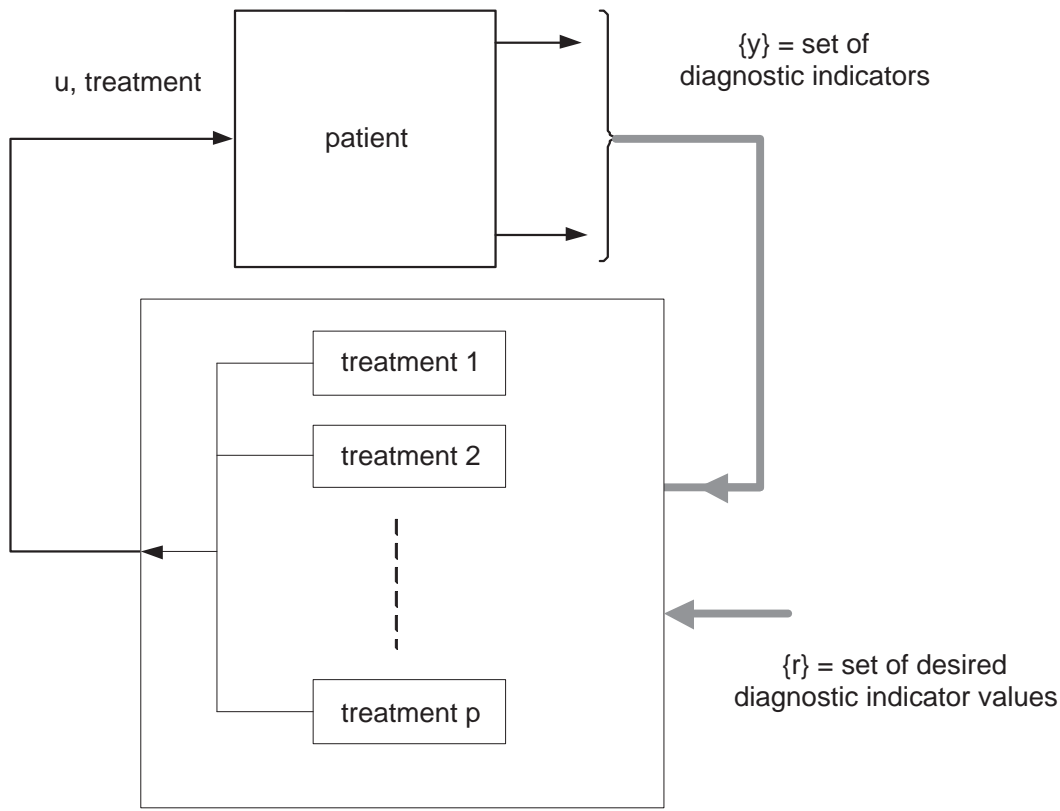}
\caption[The `switched treatment' of disease as a hybrid feedback control system]{The
`switched treatment' of disease as a hybrid feedback control system.}
\label{fig:feedbackmedicine3}
\end{center}
\end{figure}

If a selection of potentially suitable treatments are available,
then the physician/treatment controller can try them in turn with
the aim of finding a treatment that works. Switching treatments in
this way creates a \emph{hybrid control system} in which the
physician changes between a set of $p$ fixed controllers (each one
corresponding to a particular drug) as shown in Figure
\ref{fig:feedbackmedicine3}. The crucial point here is that hybrid
(or switched mode) control systems are known to be difficult to
design and analyse \cite{Pettit95}. In particular, switching
between control strategies that are themselves stable can easily
lead to an unstable situation. This is again associated with the
transient components of a controller -- essentially switching
between treatments before the transient elements of the first
treatment have gone -- causing unexpected behaviour. As an
everyday example, the control systems which regulate ABS braking
systems are hybrid controls with a very large number of switching
rules, most of which are selected on a trial and error basis to
give a stable braking control algorithm \cite{Pettit97}. An
automotive braking system is indescribably more simple than the
human body. Thus it is to be expected that physicians will have
difficulty deciding the efficacy of a particular treatment if they
switch treatments before the effects of the previous treatments
have disappeared.

\subsection{Sampling Issues in Diagnostic Indicators}

From a dynamic systems viewpoint, any control based on samples of
a variable should pay close attention to errors inherent to the
sampling process. One source of such errors is the limitations of
the sensing instruments themselves. Another source of errors could
be temporal variations in the indicator under study. In
particular, if there are cyclic variations in an indicator, then
taking a sample at one particular time instant may not give a
representative view of the average behaviour of the indicator. In
control systems this is closely related to the phenomena known as
`aliasing' and considerable attention is paid to understanding the
sampling process and limiting the errors introduced by it
\cite{BendatPiersol86}. Likewise, in control systems where the
feedback is applied on the basis of time sampled information, then
issues of `intersample ripple' cannot be ignored in assessing the
overall system performance \cite{MiddletonGoodwin90}. In plain
terms, we apply medical treatment on the basis of spot samples in
time, but this may not properly control the patient outputs at
times between samples.

\subsection{Summary}
In the preceding remarks we have used feedback control analogies
to indicate potential limitations of conventional medicines.
Enumerated they are:
\begin{enumerate}

\item Any medication  will have a transient effect as well as the
desired steady--state curative effect. By rapidly introducing
medication into a patient unwanted (transient) side effects may be
caused which mask a long term (e.g steady state) benefit.

\item Many if not most treatments are based upon pre--formulated
drugs, which in fact introduce restrictions on the physician's
ability to tailor a treatment to the needs of an individual --
this is the very antithesis of personalised medicine.

\item Many if not most treatments strongly influence body
components other than their target. This strong multivariable
component mitigates against the use of many drugs, whereas if
account were taken of the dynamics of interaction, and the
variations between individuals, it may be possible to handle the
interaction in a constructive way using a multivariable control
paradigm.

\item Treatment is usually administered in a sampled data format
by making diagnostic measurements and decisions at discrete times
which may be unrelated to the dynamical response rates of the
disease and treatment process. The intersample behaviour
associated with slow sampling may prevent or hamper the curative
influence of correct treatments.

\end{enumerate}

\section{Personalised Feedback Medicine}\label{section:Personalised FM}
Now let us return to Figure \ref{fig:feedbackmedicine2}, and
reconsider it as a model for personalised medicine with the aim of
makings some points about the rates at which treatments are
conducted. This is relevant to the remarks in \cite{VanderGreef05}
and elsewhere on lessons that may be learnt from traditional
medicine and the progressive introduction of treatments.

In the analysis given here, we make the simplifying assumption
that the practitioner has made an apriori diagnosis \footnote{In
our reading of feedback medicine the process of diagnosis is
intimately associated with the concept of state and state
observer/estimator theory. We will develop this idea elsewhere in
the context of maximising the possibility of correct diagnoses.}
and selected a sufficient set of ingredients for the treatment and
the initial ratio of mixing these ingredients. In the personalised
feedback medicine framework, these set the initial conditions for
the following time control sequence, repeated at each
consultation:
\begin{enumerate}

\item Compare the outputs \{\textbf{y}\} with desired state
\{\textbf{r}\} to form and a set of errors \{\textbf{e}\}.

\item Use the error set to make an set of adjustments to the
weights (amounts) of each component of the treatment.

\item Wait for next consultation and loop back to 1.

\end{enumerate}

As noted in the previous section, to reduce the number of
interventions required and to minimise the recovery time, a
physician might well prescribe a treatment designed in to correct
a disorder in one treatment session\footnote{In other words, the
drug dosage prescribed is immediately raised to the amount
calculated to combat the disorder.}. This approach is close to
what is termed a `one--step ahead' control policy in feedback
control systems. One--step ahead feedback policies are known to
have some drawbacks, including the fact that they demand fairly
`aggressive' control actions, with consequent side effects due to
multivariable interactions.

\begin{figure}[htbp]
\begin{center}
\includegraphics[width=5.5in]{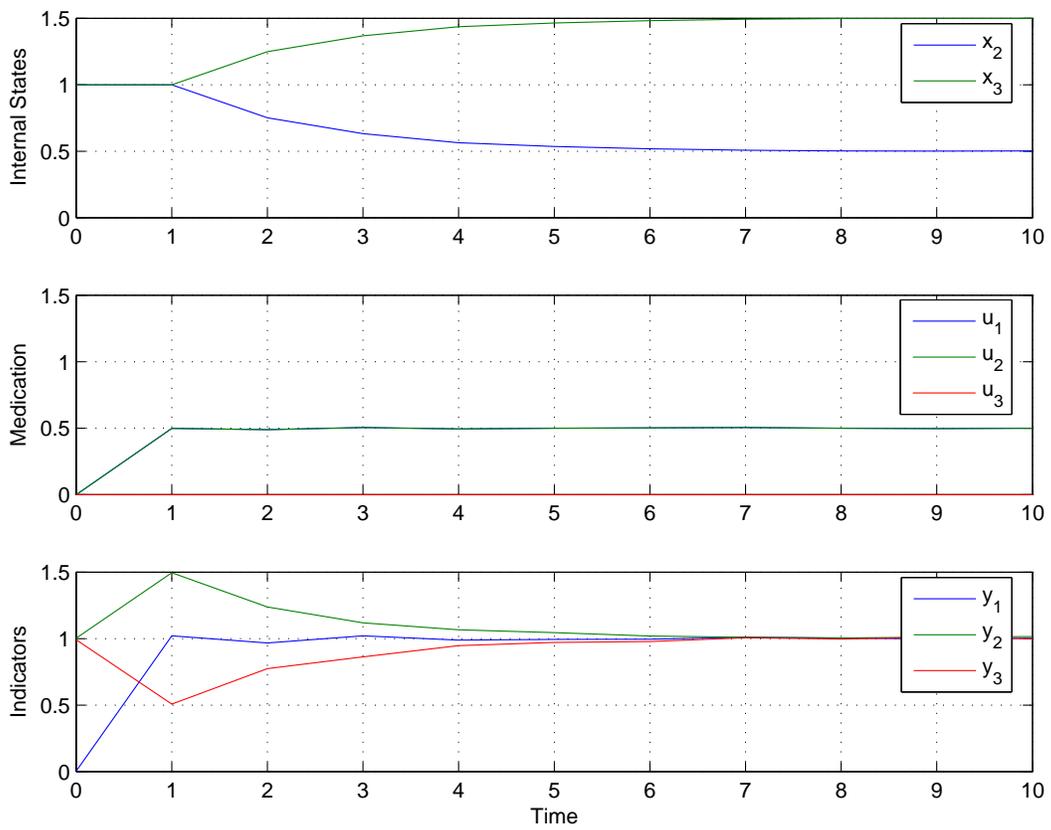}
\caption{Simulated multivariable response with aggressive feedback
control design.} \label{fig:HighGainControl}
\end{center}
\end{figure}

These multivariable interactions can be illustrated in Figure
\ref{fig:HighGainControl} which is based on a system described in
more detail later in Section \ref{sec:SideEffects}. Basically, in
this figure, a disease affecting the first output, $y_1$, is
treated by an aggressive (high gain) feedback scenario using a
combination of two independent drugs, $u_1$ and $u_2$ to return
this variable to homeostasis. The drugs cause interactions with
other variables, $y_2$ and $y_3$ which are returned to homeostasis
by the body's internal regulatory feedback loops (via the
variables $x_2$ and $x_3$). However, there are substantial, and
potentially undesirable, transients in the variables $y_2$ and
$y_3$ of approximately 50\%.

Alternatively, the adjustments can be made a little more slowly to
allow the body to adapt to the potentially damaging effects of the
transient interactions and in a way that is consistent with the
control theoretic ideas mentioned earlier. Specifically, in the
`slow change' of treatment theory, the metabolism adapts to cancel
the influence of potential toxic elements of the treatment. (This
is illustrated later in Figure \ref{fig:SIMOControlInexactSP}
where the transient side effects are reduced to about 15\% albeit
at the cost of an approximate doubling of the time before complete
regulation is achieved.)

Likewise, an additional consideration could be that the slow
change in medication will avoid de--stabilising the complex
interacting multivariable system that is the human body.
Essentially, in a feedback loop the dynamics of the body in
processing medication interact with the dynamics of the treatment
regime in ways that can be de--stabilising. Thus traditional
medicine may have, in one of its aspects, empirically discovered
ways of designing a stabilising feedback controller. We explore
this possibility and others in the following sections using
simplified models of the disease process.

Thus we would argue that there is a trade--off in the treatment
prescribed between on the one hand minimising recovery times (and
simultaneously reducing the number of consultations required)
versus reducing transient interactions and their associated
problems with transient side effects and potential
destabilisation.

The idea of feedback medicine is to bring control systems theory to health be control of
`exogenous' signals (drugs and treatments etc). The internal feedbacks associated with
Homeostatis and Allostatsis can be referred to as `endogenous' control. What the example
demonstrated is that the exogenously introduced controllers can ``fight'' the endogenous
controls.

\section{Coordinated Treatment: An Illustrative Example}

Consider a framework in which we have two different drugs
available, A and B, for the treatment of a single disease. We also
suppose that on average, across the population, both drugs have a
similar affect on the disease. Also suppose that the disease has
been correctly diagnosed with a single diagnostic indicator to be
used for monitoring and regulation purposes. There are therefore
three main strategies for treatment: (i) use drug A (only); (ii)
use drug B (only); or (iii) use a combination of both drugs A and
B. This latter strategy we will refer to as `Coordinated
Treatment' -- which is equivalent to the combinatorial treatment
terminology of \cite{VanderGreef05}. This coordinated treatment
option, although more complicated to consider, does seem to offer
some advantages that we illustrate by a simple \emph{in silico}
example.

To simplify the example, we consider one main state variable, $x$
which can be directly measured. We denote the value of the state
at each test time (for example weekly tests), $k=0,1,2...$ by
$x_k$, and the drug input(s) at each time by ${u_A}_k$ and
${u_B}_k$. If for a given patient, the two drugs, A and B, have
sensitivities $s_A$ and $s_B$ (respectively)\footnote{Here we
define sensitivity to be the response size per unit quantity of a
drug.}, then we describe a very simple\footnote{In this example,
we are assuming simple linear independent effects from the drugs,
and that the natural residence time of the state is much smaller
than the sampling time.} form of patient behaviour by:
\begin{equation}\label{eq:BasicPatientDynamics}
    x_{k+1}=s_A \times {u_A}_k + s_B \times {u_B}_k
\end{equation}

The target value for $x$ is taken to be 100\%, but we assume that
the patient is in a diseased state wherein without intervention,
$x$ becomes zero within one time interval. Therefore, some form of
drug intervention is needed and the treatment should be designed
to at least return $x_k$ to within 50\% of the target value. We
shall use the symbol $e_k$ to denote the deviation of $x_k$ from
the target value, that is $e_k=x_k-1$.

\subsection{Treatment Regimes}

We consider two main treatment regimes. In each case, the
physician monitors, at (say) a weekly interval, the indicator,
until either: (a) the physician makes a decision to discontinue
treatment (for example if there is a negative response to the
treatment); or (b) the indicator is within +50\% to +150\% in
which case the therapy is deemed to have succeeded, and the
current drug dosage is maintained.

Until a treatment regime is completed, the physician changes the
drug dosage by an amount proportional\footnote{Note that here we
ignore learning and adaptation in the physician's behaviour. This
would be harder to model, but may lead to improved treatment
completion times.} to the error between the actual, and target
indicator value (100\%). In this example, we take the
proportionality as 1/2, which corresponds to a cautious approach
to treatment, trying to avoid (if possible) over medication.

\subsubsection{Single Drug Prescription}
Suppose that the treatment regime is the use of a single drug,
(since the problem is symmetric we chose A), which is adjusted
according to the rule that the drug input is decreased by an
amount proportional (in this case 50\%) to the deviation of the
indicator from the target. This can be described by the rule:
\begin{eqnarray}
\label{eq:SingleDrugInput}
  uA_k &=& uA_{k-1}-0.5 \times (x_k -1) \\
  uB_k &=& 0 \nonumber
\end{eqnarray}

Combining equations (\ref{eq:BasicPatientDynamics}) and
(\ref{eq:SingleDrugInput}) with a little algebra leads to:
\begin{equation}\label{eq:1DrugClosedLoop}
    e_{k+1}=\left(1-\frac{s_A}{2}\right)e_k
\end{equation}

\subsubsection{Combined Drug Prescription}\label{subsubsec:CombinedDrugTreatment}
The second treatment regime we consider is the use of a equal
doses of both Drug A and Drug B. The total dosage is adjusted
according to the rule that the total drug input is decreased by an
amount proportional (in this case 50\%) to the deviation of the
indicator from the target. This can be described by the rule:
\begin{eqnarray}
\label{eq:CombinedDrugInput}
  uA_k+uB_k &=& uA_{k-1}+uB_{k-1}-0.5 \times e_k \\
  uB_k &=& uA_k \nonumber
\end{eqnarray}

Combining equations (\ref{eq:BasicPatientDynamics}) and
(\ref{eq:CombinedDrugInput}) with a little algebra leads to:
\begin{equation}\label{eq:2DrugClosedLoop}
    e_{k+1}=\left(1-\frac{s_A+s_B}{4}\right)e_k
\end{equation}

We now turn to consider simple measures of the probability of
treatment success for variable patient responses to the drugs.

\subsection{Treatment Success for Variable Patients}

Of course, if an individual patient responds as expected to the
drugs according to the average of the population $s_A=s_B=1.0$,
treatment will be `successful' in a single step, whereby the
indicator is (just) within the acceptable range. However, if we
allow for inevitable variability in an individual patient's
sensitivity to the drugs, things are slightly more complicated. In
fact, for some patients, the treatment may not lead to a
successful outcome at all, namely, where the treatment strategy
chosen will never lead to a situation where the indicator is
within the target range.

The time update equation (\ref{eq:1DrugClosedLoop}) shows that
provided $\left|1-\frac{s_A}{2}\right|<1$ at each time step, the
size of $e_k$ will decrease (exponentially) and therefore the
treatment will eventually be successful (at least in sufficient
time). On the other hand, if $\left|1-\frac{s_A}{2}\right|>1$ the
treatment regime will cause the indicator to diverge from the
target value. The main reason that this might occur is if $s_A<0$,
that is the selected drug A has a negative impact on the
particular patient. Note that it is also possible that divergence
could occur if $s_A$ is positive and much larger than expected,
i.e. hyper sensitivity in the patient. For simplicity, we ignore
this case since this is not a practical scenario since: (i) the
physician would almost certainly notice, in the response of the
indicator, the patient's hyper-sensitivity, and therefore change
the treatment strategy; (ii) If the strategy were not changed,
then it can be shown that within a few time steps, negative
amounts of drug would be prescribed. This is clearly impractical
and the simple model in (\ref{eq:SingleDrugInput}) would break
down.

Under these assumptions, it is clear that provided $s_A>0$ then
eventually, treatment with drug A only will succeed. Similarly,
from (\ref{eq:1DrugClosedLoop}) it can be shown that treatment
will succeed within $k$ steps provided
$s_A>2\left(1-2^{-1/k}\right)$. So for example, if we wish to
examine when successful treatment can be expected within $k=4$
time periods, we need $s_A>0.3182$.

Note that in the case of the coordinated treatment, the situation
is the same except that in all cases now, we replace the single
sensitivity to an individual drug $s_A$ by the patient's average
sensitivity to the two drugs $\frac{1}{2}(s_A+s_B)$. For example
therefore, the coordinated treatment will be eventually successful
if $s_A+s_B>0$ and the treatment will be successful within $n=4$
time periods if $s_A+s_B>0.6364$.

To illustrate the potential benefits of the drug treatment regime
of Subsection \ref{subsubsec:CombinedDrugTreatment} we perform
some simulations and analysis on a simplified scenario. Suppose
that the patients' sensitivities to either drug A or drug B are
independent of each other, and that in each case, 95\% of patients
will respond positively to the drug. If we assume a normal
distribution for the patient sensitivities, then both $s_A$ and
$s_B$ will have mean value of 1 and standard deviation of 0.6.
However, under the assumption that they are independent, the
distribution of the average sensitivities is also a normal
distribution with mean 1, and standard deviation
$0.6/\sqrt{2}=0.424$. This somewhat modest decrease in variability
give a considerable reduction in failure rates for the treatment.
In fact, the 5\% failure rate for a single drug, under this model,
would reduce to 1\%. Similarly, the rate of failures within four
time cycles would reduce from 13\% for the single drug treatment
to 5\% for the combined drug treatment.

\begin{figure}[htbp]
\begin{center}
\includegraphics[width=5.5in]{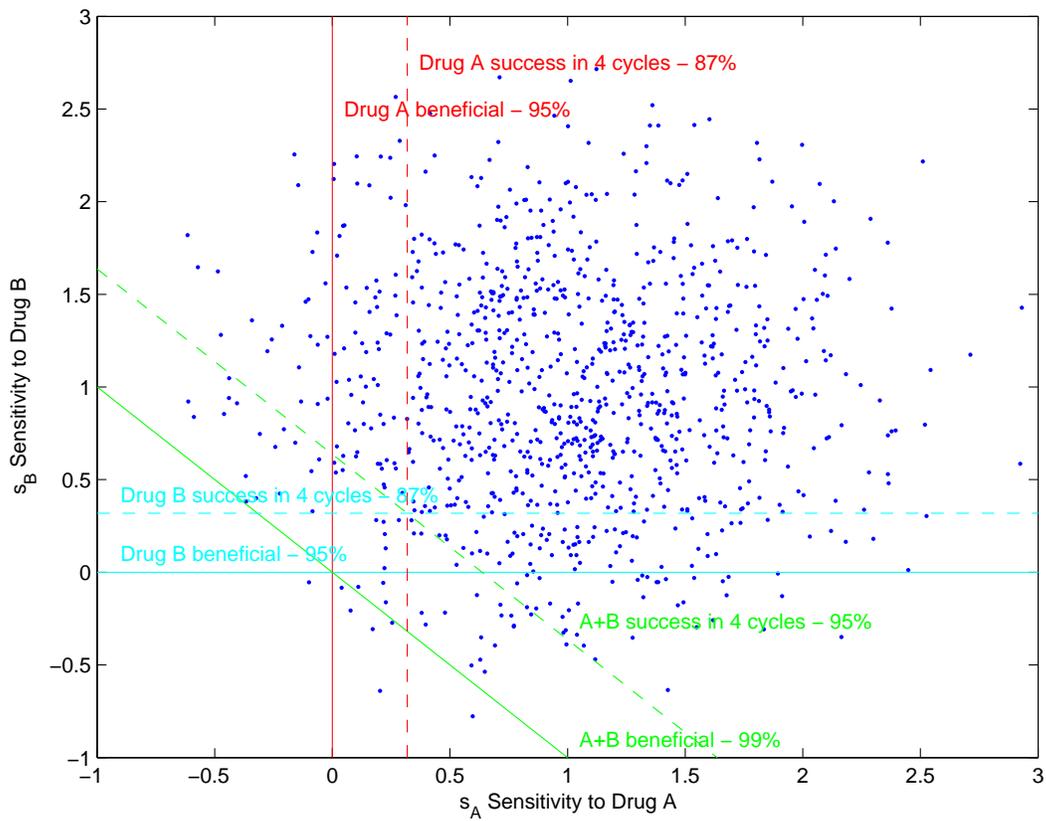}
\caption{Simulated patient sensitivities for two equally effective drugs, A and B,
together with coordinated treatment $\frac{1}{2}A+\frac{1}{2}B$.}
\label{fig:drugsensitivities}
\end{center}
\end{figure}

Figure \ref{fig:drugsensitivities} illustrates this behaviour by
showing the combination of sensitivities based on a random
selection of 1000 patients according to the distributions
described above. It also shows the dividing lines of where
different treatment regimes will be effective or not, and where
boundaries of where the different treatment regimes could be
expected to complete within four time cycles.

\subsection{Side Effects}
Another advantage of the treatment regime suggested in Subsection
\ref{subsubsec:CombinedDrugTreatment} that might be predicted by a
dynamic systems approach is a reduction, in some sense, of the
side effects of the medication. In the simplest case, we ignore
dynamic affects and patient variability, and look at the treatment
using different combinations of drugs. So in the ideal case, where
$s_a=s_B=1$, then we can achieve the desired response at the
indicator of $x=1$ by taking any combination of drugs that
satisfies:
\begin{equation}
\label{eq:allPossibleControls} u_A+u_B=1; u_A \ge 0; u_B \ge 0
\end{equation}

There are a number of different ways of measuring the overall side
effects due to independent drugs. The simplest
measure\footnote{From a mathematical viewpoint, the measures
proposed are different `norms' that might be used to quantify the
overall size of the drug dose vector $[u_A,u_B]$.} ($m$) might be
the total drug dose: $m_1=u_A+u_B$. We would argue that a more
realistic measure might be either the peak drug dose
$m_\infty=max\{u_A,u_B\}$ or the root mean square drug dose
$m_2=\sqrt{u_A^2+u_B^2}$. Both of these measures better capture
the qualitative features of many side affects whereby small
deviations can be compensated for by the body's natural regulatory
systems, but where larger deviations may exceed the body's ability
to reject this side effect.

Note that in the example we are considering, as far as side
effects are concerned:
\begin{itemize}
    \item {For any of the side effect measures $m_1$, $m_2$ or $m_\infty$ no strategy that achieves the required
    regulation, (\ref{eq:allPossibleControls}), does better than the coordinated treatment $u_A=u_B=0.5$}
    \item {For either of the side effect measures $m_2$ or $m_\infty$ the strategy that `minimises' side effects whilst
    achieving regulation,(\ref{eq:allPossibleControls}), is the coordinated treatment $u_A=u_B=0.5$ }
\end{itemize}

Note that this strategy can be thought of as application of the
feedback control concept of controller direction `alignment' (see
for example \cite{FreudenbergMiddleton99} where this is discussed
for the dual problem of single input two output systems). Broadly
speaking, this concept can be interpreted as saying that all other
things being equal, to minimise the required control responses,
and also to improve robustness, the relative weightings on the
combination of treatment components (i.e. inputs) used should
approximately reflect the responsiveness of the patient to each
individual treatment component. It is also known from the control
literature that this will in general produce a more robust control
algorithm than strategies that don't respect this suggestion.

\section{Side Effects and (partial) Homeostasis}\label{sec:SideEffects}

In this section, we wish to illustrate some side effects that can
occur when trying to regulate a multivariable system. We consider
the simple case of a system (modelling in a crude way homeostasis)
with 3 important variables, each of which we assume can be
measured. We denote these variables by $y_1$, $y_2$ and $y_3$
respectively. Each of these variables has two main influences - a
contribution due to external inputs (i.e. medication) and internal
biological regulator processes. The internal regulatory processes
have state variables that we denote by $x_1$, $x_2$ and $x_3$
respectively. We also suppose that the three drug components
available, $u_1$, $u_2$ and $u_3$ have interactions between the
different variables given by the matrix $M_u$. We then take the
simple (static) mathematical model:

\begin{equation}
\label{eq:MIMOegPlantInteractions}
\left[ \begin{array}{c}
  y_1 \\
  y_2 \\
  y_3 \\
\end{array}\right] = M_u \times
\left[\begin{array}{c}
  u_1 \\
  u_2 \\
  u_3 \\
\end{array} \right]+
\left[\begin{array}{c}
  x_1 \\
  x_2 \\
  x_3 \\
\end{array} \right]
\end{equation}

We take, for illustration,
$M_u=\left[
\begin{array}{ccc}
  1 & 1 & 0 \\
  1 & 0 & 1 \\
  0 & -1 & -1 \\
\end{array}%
\right]$ whereby, the first drug component has a positive effect
on indicators 1 and 2; the second drug component has a positive
effect on indicators 1 and 3; and the third drug suppresses (i.e.
has a negative effect) on indicators 2 and 3. In a healthy
individual, we assume that biological regulatory processes add
dynamics to achieve perfect adaptation through integral feedback
control \cite{Yi_etal2000}. We describe these processes by the
differential equation:

\begin{equation}
\label{eq:BioRegulation} \frac{d}{dt} \left( x_i(t) \right) =
r_{bi} - y_i(t); i=1..3
\end{equation}
where $r_{bi}$ are internal biological references for which
(\ref{eq:BioRegulation}) will generate actions through negative
feedback in an attempt to achieve Homeostasis, that is,
$y_i(t)\rightarrow r_{bi}$.

In our simple example, we take each of the biological references
to be unity. We also consider a case where disease affects the
first set of state and output variables and forces, instead of
(\ref{eq:BioRegulation}), the first state component to obey
$x_1=0$. Note that throughout this section, we constrain all
variables to take non-negative values.

We now wish to consider two different approaches to medical
intervention (external regulation) of such a patient.

\subsection{External `Perfect' Regulation of All Variables}

In the first approach to medical intervention, suppose the
physician, knowing that there are 3 drugs available, and 3
diagnostic indicators, tries to adapt and regulate all variables
to reference values that we denote by $r_{pi}:i=1..3$. Ideally,
these reference values should be equal to the internal biological
reference values, $r_{bi}:i=1..3$. Also, let us suppose that the
physician knows of the drug interactions, and the matrix $M_u$ and
therefore is able to compensate for these interactions in the
proposed treatment regime using the inverse
$M_u^{-1}=0.5\left[
\begin{array}{ccc}
  1 & 1 & 1 \\
  1 & -1 & -1 \\
  -1 & 1 & -1 \\
\end{array}%
\right]$. The treatment regime proposed\footnote{Technically, the
multiplication by $M_u^{-1}$ should be after the controller, not
before, however, because the control proposed (after compensation)
is identical in all channels, this makes no difference to the
analysis in this case.} is then described by the difference
equation\footnote{Note that this difference equation is
qualitatively very similar to the differential equation
(\ref{eq:BioRegulation}) except that in this case, the physician's
interventions can only be altered at a discrete set of times,
which are indexed by the variable `$k$' in
(\ref{eq:MIMOControl}).}:
\begin{equation}
\label{eq:MIMOControl}
\left[ \begin{array}{c}
  {u_1}_{k+1}  \\
  {u_2}_{k+1}  \\
  {u_3}_{k+1} \\
    \end{array}
  \right] =
\left[ \begin{array}{c}
  {u_1}_{k}  \\
  {u_2}_{k}  \\
  {u_3}_{k} \\
  \end{array}\right]
  + 0.5 \times M_u^{-1}
  \left[%
\begin{array}{c}
  r_{p1}-{y_1}_k \\
  r_{p2}-{y_2}_k \\
  r_{p3}-{y_3}_k \\
\end{array}%
\right]
\end{equation}

\begin{figure}[htbp]
\begin{center}
\includegraphics[width=5.5in]{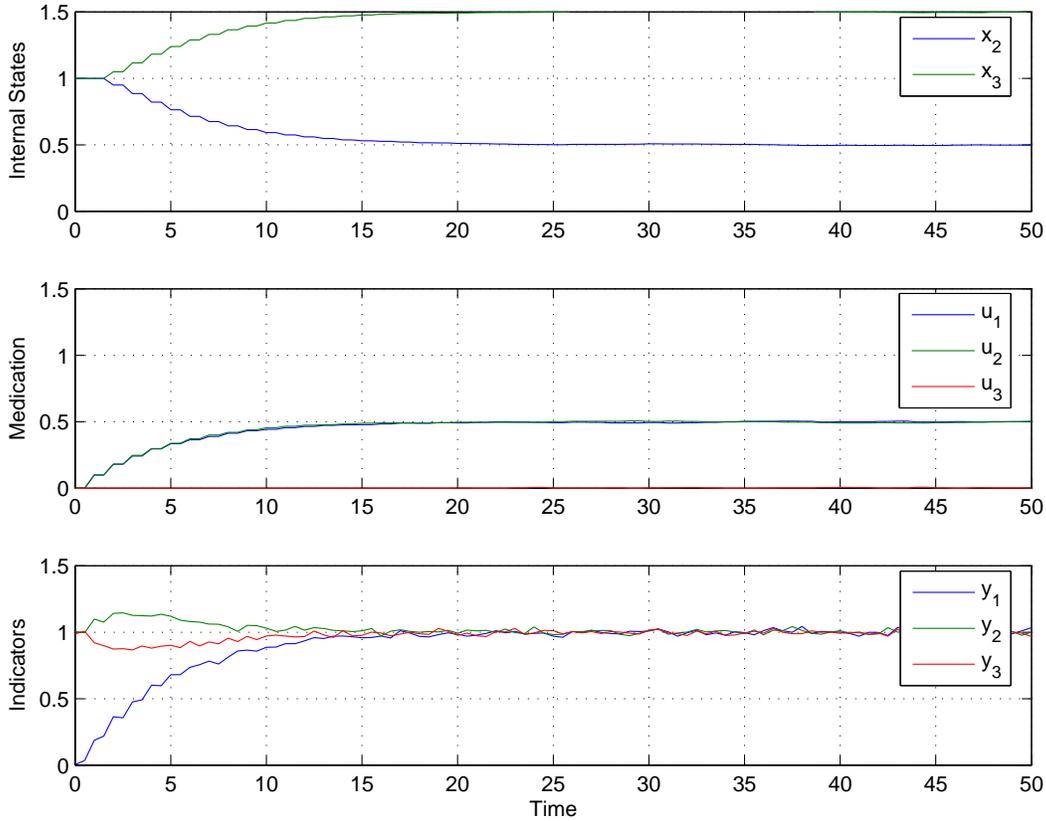}
\caption{Simulated patient transient response to treatment with
external regulation of all variables to the exact biological
values.} \label{fig:MIMOControlExactSP}
\end{center}
\end{figure}

In this situation, and in the ideal case where the biological
reference values, $r_{bi}$ are identical to the physician's
reference values $r_{pi}$ we obtain simulation results as given in
Figure \ref{fig:MIMOControlExactSP}.

In Figure \ref{fig:MIMOControlExactSP}, the responses are all well
behaved, with the medication fairly rapidly converging to values
that give all variables at their target values of $1.0$.

\begin{figure}[htbp]
\begin{center}
\includegraphics[width=5.5in]{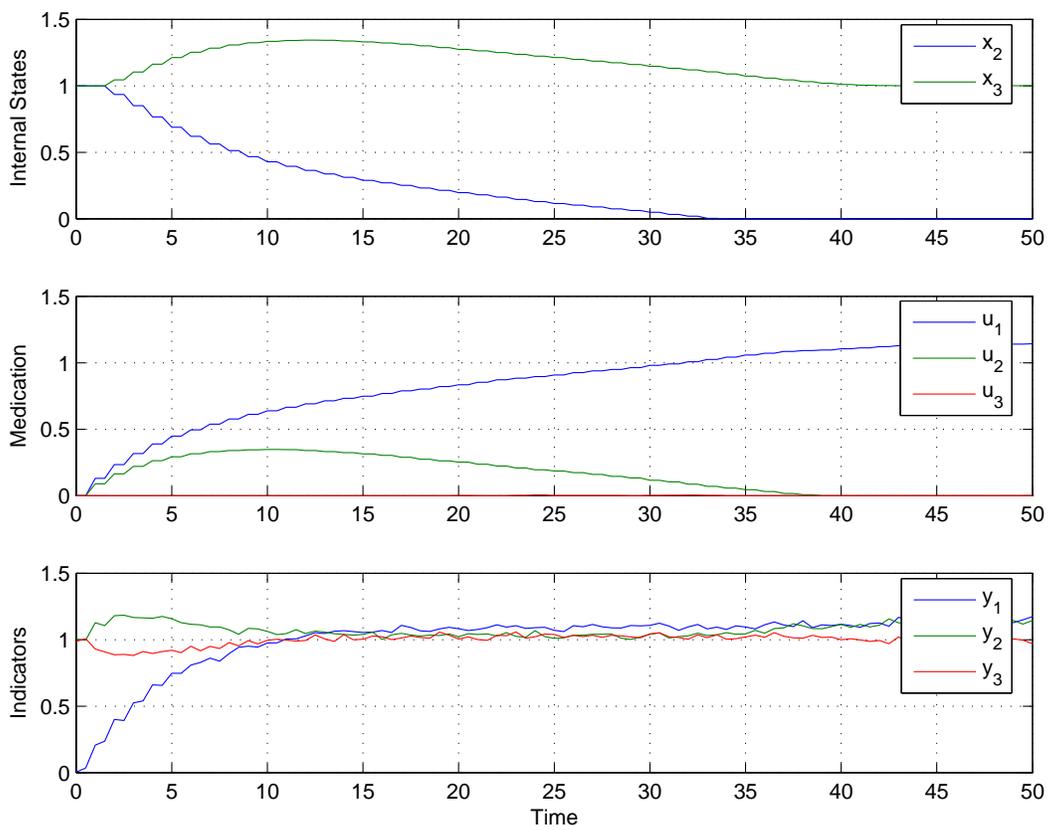}
\caption{Simulated patient transient response to treatment with
external regulation of all variables with 10\% deviation between
internal biological and external reference values.}
\label{fig:MIMOControlInexactSP}
\end{center}
\end{figure}

However, if we repeat the simulations, except in this case with a
slight (10\%) deviation between the internal biological reference
values $r_{bi}$ and the physician's external reference values
$r_{pi}$. In this case, we obtain the simulation results shown in
Figure \ref{fig:MIMOControlInexactSP}.

From this figure we see that at least initially (up to about 5 to
10 time samples), the response is quite close to that shown
earlier in Figure \ref{fig:MIMOControlExactSP}. However, after
this time, a continual drift, ending in quite substantial
deviations can be seen.

This drift can be observed in both medication levels, and internal
states of the patient, but is largely unobservable in the
behaviour of the diagnostic indicators. In this simulation
example, the drift continues until some of the regulators are so
far from normal conditions that they saturate and they themselves
are no longer capable of maintaining homeostasis. In this
particular example, the regulators that saturate are:
\begin{itemize}
    \item {Internal regulation of the 3rd state, $x_3$ which reaches
    zero at approximately the 34th sample, and thereafter is clamped at zero.}
    \item {External regulation via administration of the 2nd drug component, $u_2$,
    which reaches zero a short time later, and thereafter is clamped at zero.}
\end{itemize}

In this case, we see that the internal and external regulatory
loops are `fighting' each other, namely, they are both trying to
achieve perfect adaptation, but to slightly different reference
values. This results in the undesirable behaviour leading to much
higher than necessary drug dosages (in some components of the
drug), and saturation of internal biological regulatory
mechanisms.

From a dynamical systems point of view, it can be shown that the
patient model described above, gives a transfer function that has
two (steady state) transmission zeros, in other words, the steady
state response patient response drops from rank $3$ (as might
normally be considered for a three output system) by $2$ (due to
the transmission zeros) to a rank $1$ matrix. It is well known in
dynamical systems, that to close $m$ integral feedback loops (i.e.
perfect adaptation mechanisms) around a plant, the steady state
transfer matrix should have rank no less than $m$ or else there
will necessarily be marginally stable hidden modes (i.e. hard to
observe drifting modes) in the response. This explains the poor
behaviour exhibited in this example.

Note that this example is not an isolated case. Indeed, any time
that we have measurement of a variable for which there is an
internal regulatory loop that achieves perfect regulation, it
necessarily follows that there will be a transmission zero to this
variable, and external perfect adaptation should not normally also
be employed for this variable.

\subsection{Coordinated Regulation of a Single Variable}
An alternative approach, suggested by dynamical systems theory,
would be to allow integral action (`perfect adaptation') of a
single variable only, namely, $y_1$ for which the biological
regulatory system is inoperative.

In this case, we consider a single dynamic variable, $v_k$ which
we update by the equation:
\begin{equation}\label{eq:SIMOIntegrator}
    v_{k+1}=v_k+0.5\times \left( r_{p1}-{y_1}_k \right)
\end{equation}
(\ref{eq:SIMOIntegrator}) together with the control action
equation (suggested in view of the discussions in Section
\ref{subsubsec:CombinedDrugTreatment}):

\begin{equation}\label{eq:SIMOControlDirection}
u_k=\left[ \begin{array}{c}
  0.5 \\
  0.5 \\
  0 \\
\end{array}\right] v_k
\end{equation}

\begin{figure}[htbp]
\begin{center}
\includegraphics[width=5.5in]{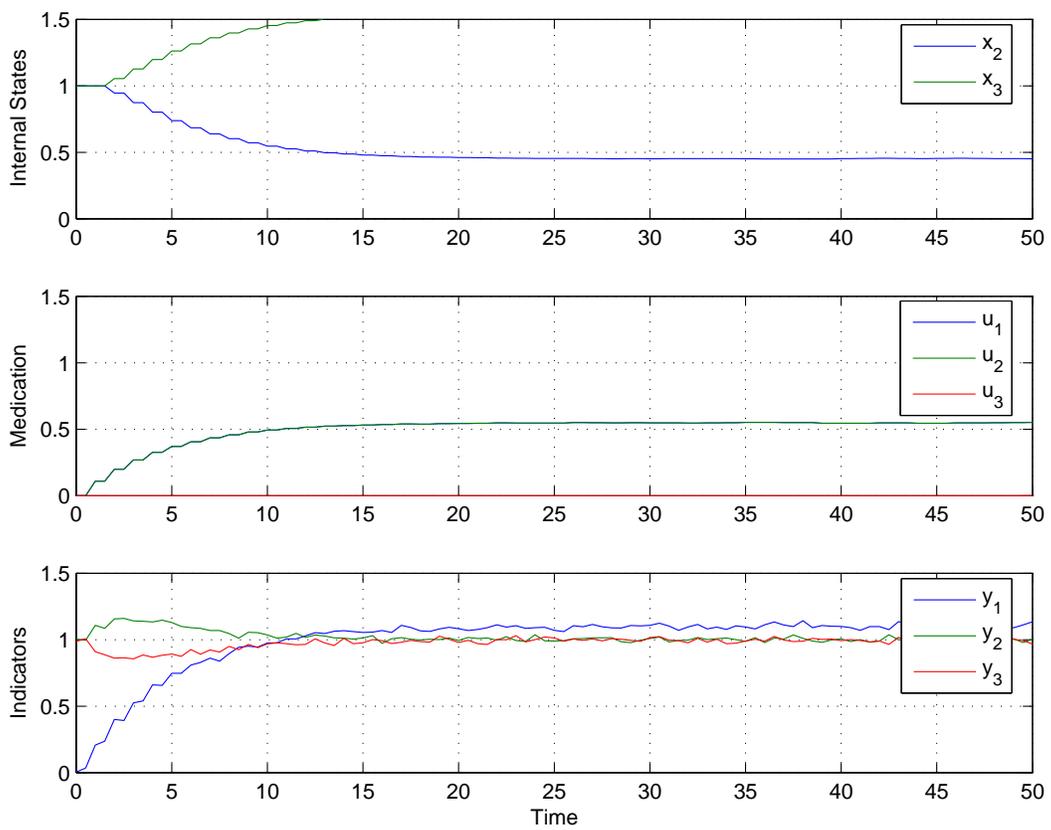}
\caption{Simulated patient transient response to treatment with
external regulation of first variable only, allowing for 10\%
error in reference values.} \label{fig:SIMOControlInexactSP}
\end{center}
\end{figure}

Using the treatment regime described by
(\ref{eq:SIMOControlDirection}), (\ref{eq:SIMOIntegrator}) (in
place of the (\ref{eq:MIMOControl})  ) we obtain the simulated
responses given in Figure \ref{fig:SIMOControlInexactSP} even with
discrepancies between the internal biological reference variable
(which in this case is deactivated) $r_{b1}$ and the external
reference $r_{p1}$.

Note that in this case the response is close to the ideal case,
with little use of drugs and without saturating any of the natural
feedback mechanisms.

\subsection{ Remarks on `Linear' Systems}

\begin{figure}
\begin{center}
\includegraphics[width=5.5in]{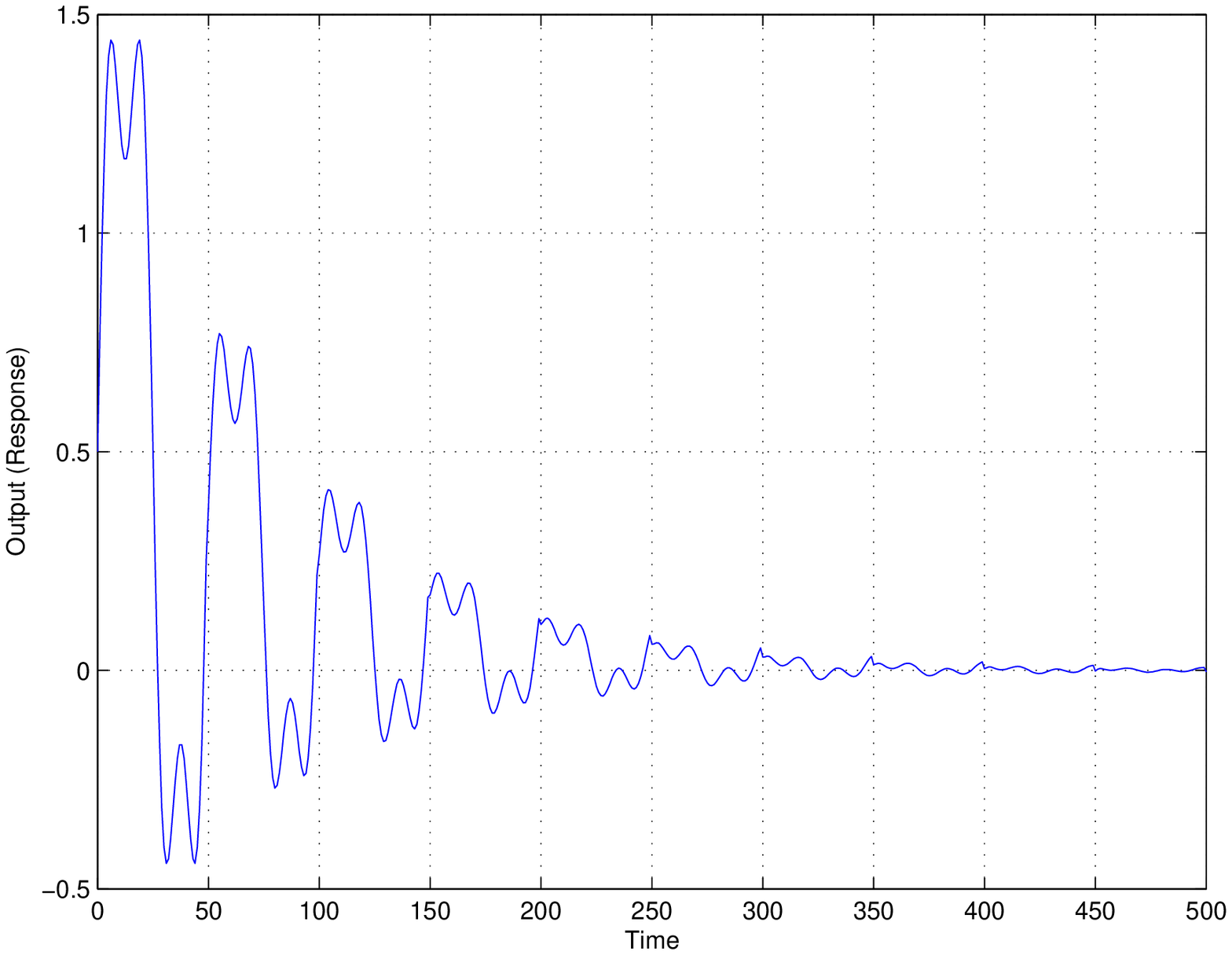}
\caption[Example of a complicated response arising from a purely
linear system.]{Example of a complicated response arising from a
purely `linear' system.} \label{fig:linear_system_eg}
\end{center}
\end{figure}

Our discussions here are in some cases based on what might be
termed \emph{independent responsiveness} in a system. In a
feedback control systems context, this is often termed
\emph{linearity}, however we believe this term has an alternate
usage in other disciplines where linearity might be taken to imply
`simple', `smooth', easily predictable behaviour. In particular,
from a feedback control systems perspective, `linear' systems
permit quite complex behaviours, as illustrated in Figure
\ref{fig:linear_system_eg}.

The feedback control systems of linearity  or \emph{independent
responsiveness} can be defined by the two key properties of:
    \textit{(i) proportionate response} -- which loosely speaking means that if an action
    produces a certain response, an increased action will produce
    an increased response\footnote{Technically, the increased response should be proportionate to the increase in action}.
    \textit{(ii) cumulative affects} -- which loosely means that the result
    of combined actions can be reasonably predicted simply
    by combining\footnote{Mathematically, the response to a sum of actions should be exactly
     the sum of the individual responses to each action.} the results of individual actions.

Whilst it is certainly true that there are heavily dependent
responses (non--linearities) in the dynamics of biomedical
systems, we argue that a study based on linearisation is an
appropriate first step. Our primary reasons for this are that the
linear case is frequently a useful approximation of a non--linear
system; and, that the theory and understanding of linear dynamic
systems is far more straightforward and developed than the
equivalent theory for non--linear systems. Furthermore, the
framework we propose is aimed at keeping homeostasis, and avoiding
large perturbations to this where nonlinear affects will be more
pronounced. A more extensive study could include extra nonlinear
features such as: (i) Saturation of variables (i.e. in several
cases, it may not be possible, or there may be unacceptable side
effects of variables exceeding certain ranges); (ii) Complicated
variable interactions (e.g.\ the use of drugs which on their own
are acceptable, but who's combined use may be contra--indicated;
or conversely, drugs which on their own have a small therapeutic
effect, but which combine in a ``strongly" synergistic manner).

\section{What Next?}
In these notes we ask whether knowledge from feedback control
theory can be employed to make meaningful statements about
personalised and combinatorial medicine. To do this we have made
assumptions and simplifications which maybe inappropriate.
Likewise or examples and conclusions may be obvious to biologists
and medical researchers. In this spirit, we invite scientific
feedback and comments on the potential relevance of the feedback
medicine concept and the examples used in this document.

\section*{Acknowledgements}
P. W. acknowledges the support of the Science Foundation Ireland
thought its Research Professor programme. R.M. acknowledges the
support of the Hamilton Institute Distinguished Visitor Fund and
the University of Newcastle, Australia. O.W. acknowledges the
support of the German Federal Ministry for Education and Research
(BMBF) as part of the SMP Protein in the National Genome Research
Network (NGFNII).

\begin{comment}

Working note  -- Peter Wellstead, first edit 30th. August 2005, Edit two 13th September
2005, third edit 22nd Sept, edit four 17 Oct. Rick Middleton first edit 20th Sept, edit
two 5 Oct. Peter Wellstead inserting Olaf's comments and additions 10th Jan 2006. Peter
Wellstead text corrections 13th January. Olaf Wolkenhauer text corrections 18th March
2006.

\end{comment}

\bibliographystyle{unsrt}
\bibliography{feedbackmedicine}
\end{document}